\documentclass[manuscript]{emulateapj}

\usepackage[]{natbib}
\usepackage{pifont}

\newcommand{\ergsec}{erg~s$^{-1}$}

\newcommand{\ergsechz}{erg~s$^{-1}$~Hz$^{-1}$}
\newcommand{\lya}{Ly$\alpha$}
\newcommand{\hi}{H{\sc i}}
\newcommand{\halpha}{H$\alpha$}
\newcommand{\hbeta}{H$\beta$}
\newcommand{\wlya}{$W_{\mathrm{Ly}\alpha}$}
\newcommand{\klya}{$k_{\mathrm{Ly}\alpha}$}
\newcommand{\clya}{$C_{\mathrm{Ly}\alpha}$}
\newcommand{\fesclya}{$f_{\mathrm{esc}}^{\mathrm{Ly}\alpha}$}

\newcommand{\fesclyc}{$f_{\mathrm{esc}}^{\mathrm{LyC}}$}
\newcommand{\rhol}{$\rho_{\mathrm{L}}$}

\newcommand{\rholuv}{$\rho_{\mathrm{L,UV}}$}
\newcommand{\rhodstar}{$\dot \rho_{\star}$}
\newcommand{\rhodstarlya}{$\dot \rho_{\star}^{\mathrm{Ly}\alpha}$}
\newcommand{\rhodstaruv}{$\dot \rho_{\star}^{\mathrm{UV}}$}

\newcommand{\ebv}{$E_{B-V}$}
\newcommand{\msunyr}{$M_\odot$~yr$^{-1}$}
\newcommand{\msunyrvol}{$M_\odot$~yr$^{-1}$~Mpc$^{-3}$}

\slugcomment{Accepted by the Astrophysical Journal}

\shorttitle{On the redshift evolution of the Lyman-alpha escape fraction }
\shortauthors{Matthew Hayes et al.}

\begin{document}

\title{On the redshift-evolution of the Lyman-alpha escape fraction and the 
		dust content of galaxies}

\author{Matthew Hayes\altaffilmark{1}, Daniel Schaerer\altaffilmark{1,2},
		G\"oran \"Ostlin\altaffilmark{3}, J. Miguel Mas-Hesse\altaffilmark{4}, Hakim
				Atek\altaffilmark{5}, Daniel Kunth\altaffilmark{6}}
\email{matthew.hayes@ast.obs-mip.fr}

\altaffiltext{1}{Observatory of Geneva, University of Geneva, 51 chemin des Maillettes, 1290 Versoix, Switzerland}
\altaffiltext{2}{Laboratoire d'Astrophysique de Toulouse-Tarbes, Universit\'e de Toulouse, CNRS, 14 Avenue E. Belin, 31400 Toulouse, France}
\altaffiltext{3}{Oskar Klein Centre, Department of Astronomy, AlbaNova
		University Center, Stockholm University, 10691 Stockholm, Sweden}
\altaffiltext{4}{Centro de Astrobiolog{\'i}a (CSIC-INTA), PO Box 78, 28691 Villanueva de la Ca{\~ n}ada, Madrid, Spain}
\altaffiltext{5}{Spitzer Science Center, Caltech, Pasadena, CA 91125, USA}
\altaffiltext{6}{Institut d'Astrophysique de Paris (IAP), 98~bis boulevard Arago, 75014 Paris, France}

\begin{abstract}
The \lya\ emission line has been proven a powerful tool by which to study 
evolving galaxies at the highest redshifts. However, in order to use \lya\ 
as a physical probe of galaxies, it becomes vital to know the \lya\ escape 
fraction (\fesclya). Unfortunately, due to the resonant nature of \lya, 
\fesclya\ may vary unpredictably and requires empirical measurement. Here 
we compile \lya\ luminosity functions between redshift z=0 and 8 and, 
combined with \halpha\ and ultraviolet data, assess how \fesclya\ evolves 
with redshift. We find a strong upwards evolution in \fesclya\ over the 
range $z=0.3-6$, which is well-fit by the power-law \fesclya$\propto (1+z)^\xi$
with $\xi=(2.57_{-0.12}^{+0.19})$. This predicts that \fesclya\ should reach 
unity at $z=11.1$. By comparing \fesclya\ and \ebv\ in individual galaxies we 
derive an empirical relationship between \fesclya\ and \ebv, which includes 
resonance scattering and can explain the redshift evolution of \fesclya\ 
between $z=0$ and 6 purely as a function of the evolution in the dust content of 
galaxies. Beyond $z\approx 6.5$, \fesclya\ drops more substantially; an effect 
attributed to either ionizing photon leakage, or an increase in the neutral gas
fraction of the intergalactic medium. While distinguishing between those two 
scenarios may be extremely challenging, by framing the problem this way we 
remove the uncertainty of the halo mass from \lya-based tests of reionization. 
We finally derive a new method by which to estimate the dust content of galaxies 
based purely upon the observed \lya\ and UV LFs. These data are characterized by
an exponential with an e-folding redshift of $\approx 3.5$.
\end{abstract}

\keywords{Galaxies: evolution --- Galaxies: high-redshift --- Galaxies: luminosity function, mass function --- Galaxies: star formation--- dark ages, reionization, first stars}

\section{Introduction}

Surveys targeting the Lyman-alpha emission line (\lya) show unique 
profitability for examining the formation and evolution of the galaxy population 
between redshift $z\approx 2$ and $\gtrsim 7$. \lya\ has been exploited by many
teams and the combined catalogues would currently include over two thousand 
entries 
\citep[e.g.][]{Venemans2002,Hu2004,vanBreukelen2005,Wang2005,Shimasaku2006,
Gronwall2007,Ouchi2008,Nilsson2009,Guaita2010,Cassata2010,Hayes2010b}.
Wherever such large samples are available, the temptation is strong to use their 
statistical power to examine as many physical properties of the galaxy 
population as possible. This, however, requires that the numbers one has at 
hand are in some way a physical reflection of those underlying properties; to 
first order the luminosity (and/or equivalent width for emission lines) must be 
related to its intrinsic value. For surveys that target the restframe
ultraviolet continuum (UV) this is simply a matter of applying a dust 
correction. However, the 
resonant nature of the \lya\ line means its radiation transport becomes an 
involved and detailed problem
\citep{Osterbrock1962,Adams1972,Harrington1973,Neufeld1990,Ahn2003,
Verhamme2006,Tasitsiomi2006,Laursen2009}.
This further implies that the escaping fraction of photons (\fesclya) may not 
be assumed, is liable to evolve strongly with an evolving galaxy population,
and must be measured empirically. Pursuing this line of inquiry, the evolution
of \fesclya\ can therefore provide us with independent estimates of how various
properties of the galaxy population evolve over cosmic time.

Since \lya\ photons scatter in neutral hydrogen (\hi) until they either 
escape or are absorbed by dust grains, most fundamentally the radiation 
transport depends upon the \hi\ content, 
its geometry and kinematics, and the dust content and distribution. Regrettably,
with current observational facilities, the only one of these quantities 
that can easily be estimated for large samples of high-redshift galaxies is the 
dust attenuation, which is typically derived from the stellar continuum. 
Consequently the amalgamated effects of the remaining 
quantities, and how they affect \fesclya, can only be assessed on a statistical 
basis. 

\lya\ surveys have been fruitful over the last decade, but it is only very 
recently that robust \fesclya\ measurements have been made on statistically 
meaningful samples \citep{Verhamme2008,Atek2009,Kornei2010,Hayes2010b}. 
However at the current juncture, all of these studies estimate \fesclya\ by 
different methods, and are derived among samples compiled at various redshifts 
and filtered through differing selection functions. 
Thus synthesis of the results remains somewhat difficult. 
Furthermore, there is no self-consistent study in the 
current literature of how \fesclya\ evolves with redshift and it is this point
that we take the first steps towards rectifying with the current article. 
We begin by compiling various \lya, \halpha, and UV datasets in 
\S~\ref{sect:fescz}, which we use to estimate the redshift evolution of 
\fesclya. We discuss the general trends and draw comparisons with other 
observational and theoretical methods in \S~\ref{sect:genres}. In 
\S~\ref{sect:fescebv} we investigate the effect of the one quantity that is
relatively easy to measure -- the dust content -- and discuss how it affects
\fesclya. In \S~\ref{sect:disc} we discuss the trends with redshift in more 
detail and synthesize information from \S~\ref{sect:genres} and 
\S~\ref{sect:fescebv} in order to make more 
detailed inferences about the evolution of the properties of the interstellar 
medium (ISM) of galaxies, the intergalactic medium (IGM), and the overall
dust content. 	
In \S~\ref{sect:summ} we present a final summary.
All data are scaled to a cosmology of 
$(H_0,\Omega_\mathrm{M},\Omega_\Lambda)=(70~\mathrm{km~s}^{-1}~\mathrm{Mpc}^{-1},0.3,0.7)$.

\vspace{5mm}

\section{Method: the \lya\ escape fraction measurements}\label{sect:fescz}

\subsection{Escape fraction calculations}

We now proceed to compile various estimates of \fesclya\ as a function of 
redshift, but first we present the formalism. 
We continue with the \citet{Hayes2010b} definition of \fesclya: the 
sample-averaged, ``volumetric" escape fraction. This quantity is defined as the
ratio of observed to intrinsic \lya\ luminosity densities (\rhol), derived by
integration over luminosity functions (LF), as in Equation~\ref{eq:fesc_def}:
\begin{equation}
f_{\mathrm{esc}}^{\mathrm{Ly}\alpha} = 
\frac{\rho_{\mathrm{L,Ly}\alpha}^{\mathrm{Obs}}}{\rho_{\mathrm{L,Ly}\alpha}^{\mathrm{Int}}} =
\frac{\int_{L_\mathrm{lo}}^\infty \Phi(L)_{\mathrm{Ly}\alpha}^{\mathrm{Obs}} \cdot L \cdot \mathrm{d}L}
{\int_{L_\mathrm{lo}}^\infty \Phi(L)_{\mathrm{Ly}\alpha}^{\mathrm{Int}} \cdot L \cdot \mathrm{d}L}
\label{eq:fesc_def}
\end{equation}
where $\Phi(L)$ are the standard luminosity functions\footnote{LFs are typically
parameterized by the \citet{Schechter1976} function: $\Phi(L) \cdot \mathrm{d}L =
		\phi_\star \cdot (L/L_\star)^{\alpha} \cdot \exp (L/L_\star) \cdot \mathrm{d}L/L_\star$.}. 
Thus \fesclya\ is not simply a re-scaling of the LF by $L$ (constantly scaling
the escape fraction of all galaxies) or by $\Phi$ (the duty cycle; see 
\citealt{Nagamine2008} for examples of both of these methods). Instead, since 
\fesclya\ is simply defined as the ratio of luminosity densities, it 
can be thought of as the
fraction of \lya\ photons that escape from the survey volume, regardless
of whether all galaxies show low \fesclya, or whether only a fraction of 
galaxies are in the \lya\ emitting phase with high \fesclya\ (see arguments in 
\citealt{Tilvi2009}).
By definition \fesclya\ also includes any possible effect that the IGM may have on
the \lya\ emission from galaxies. However, it is clear that the bulk of the 
evolution of \fesclya\ with redshift found in this article can clearly not be 
attributed to variations of the IGM transmission. 

Where possible (i.e. $z<2.3$) we make a direct comparison between \lya\ 
and \halpha. We apply the most appropriate dust correction to \halpha\
and multiply by the case B recombination ratio of \lya/\halpha=8.7 
\citep{Brocklehurst1971} in order to obtain the intrinsic \lya. I.e. 
\begin{equation}
f_{\mathrm{esc}}^{\mathrm{Ly}\alpha} (z<2.3)  = 
\frac{\rho_{\mathrm{L,Ly}\alpha}^{\mathrm{Obs}}}{8.7 \cdot \rho_{\mathrm{L,H}\alpha}^{\mathrm{Int}}} =
\frac{\rho_{\mathrm{L,Ly}\alpha}^{\mathrm{Obs}}}{8.7 \cdot 10^{0.4 E_{B-V} k_{6563} } \cdot \rho_{\mathrm{L,H}\alpha}^{\mathrm{Obs}}},
\label{eq:fesc_def_ha}
\end{equation}
where \ebv\ must be the dust attenuation computed for the \halpha\ 
emitting sample, and $k_{6563}$ the extinction coefficient at the wavelength of
\halpha. Superscripts Int and Obs refer to the intrinsic and observed 
quantities. 

At $z\gtrsim 2.3$ we are unable to obtain \halpha\ LFs in order to use line 
ratios to estimate \fesclya\ and instead the estimate is derived from the UV 
continuum. This is a less elegant method since the conversion between UV and 
\lya\ requires the assumption of a metallicity, initial mass function (IMF), 
and evolutionary stage. 
However, in light of the fact that higher-redshift \halpha\ studies will 
remain impossible until the arrival of the {\em James Webb Space Telescope}, 
this is the only way to proceed.
It is fortunate that there is no evidence that IMFs should differ between
\lya- and UV selected populations, although metallicities have been shown to be
around 0.2~dex lower \citep[e.g.][]{Cowie2010} which translates into a 
difference of $\lesssim 20$~\% in the intrinsic \lya/UV ratio
\citep{Leitherer1999}. For ``normal" metallicities and IMFs, and assuming that 
on average star-formation is ongoing at equilibrium, this method is the same as
taking the ratio of \lya/UV star-formation rate densities (\rhodstar): 
\begin{equation}
f_{\mathrm{esc}}^{\mathrm{Ly}\alpha}(z>2.3) = 
\frac{\dot \rho_{\star, \mathrm{Ly}\alpha}^{\mathrm{Obs}}}{\dot \rho_{\star}^{\mathrm{Int}}} =
\frac{\dot \rho_{\star,\mathrm{Ly}\alpha}^{\mathrm{Obs}}}{10^{0.4 E_{B-V} k_{\mathrm{UV}} } \cdot \dot \rho_{\star,\mathrm{UV}}^{\mathrm{Obs}}},
\label{eq:fesc_def_uv}
\end{equation}
where now \ebv\ must be the extinction seen by the UV-selected population and
$k_{\mathrm{UV}}$ is the extinction coefficient in the UV.

The UV is of course not the only wavelength we can use for this
experiment, but we choose to work exclusively with UV LFs 
since they (a) are so abundant in the literature, (b)	have 
reasonably well-understood selection functions, and (c) span an appropriately 
large range in redshift. 
We adopt UV measurements at redshifts most appropriate to our compiled \lya\ 
data and dust attenuations derived from these samples themselves. 
We further adopt the dust attenuation law of \citet{Calzetti2000}, and the
SFR calibrations of \citet{Kennicutt1998}. These calibrations assume a stabilized
star formation episode at a constant rate for longer than around 100~Myr, with a
Salpeter Initial Mass function (mass limits between 0.1 and 100 $M_\odot$), and 
a complete ionization efficiency (no leaking and no destruction of
ionizing photons by dust).
In general we assume that `UV' refers to the restframe wavelength of 1500~\AA, 
where the extinction coefficient computed from the relationship of 
\citet{Calzetti2000} is 10.3.
We want to emphasize that the definition of \fesclya\ we are using for high redshift 
galaxies includes any effect that would decrease the number of observed \lya\ photons 
with respect to the number expected from the star formation rate derived from the UV 
continuum level. The leaking of ionizing photons, as we will discuss later, would 
therefore imply an \fesclya\ value below unity, even if 100\% of the \lya\ photons 
effectively produced in the galaxy are able to escape without being affected by resonant 
trapping or destruction by dust.

\subsection{Limits of integration}\label{sect:intlims}

The goal of this study is to determine the total, volumetric escape fraction of
a given volume, and ideally would include the very faintest systems.  In 
practice this would require integration of the LFs down to zero, which depending
on the observational limits of a given survey and the redshift-dependent values 
of both $L_\star$ and $\alpha$, may include large extrapolations (or may even be
divergent). It is vital
therefore, that our study employs lower integration limits that are: (a)
self-consistent between the populations; (b) include a sufficiently meaningful
fraction of \rhol, and (c) are not dominated by over-extrapolation and 
uncertainties in the faint-end slope. 

At $z=2,3$ and $>4$, several studies of the \rholuv\ have been published,
and here we adopt those of \citet{Reddy2008} and \citet{Bouwens2009a}, 
respectively. Both perform integrations down to $0.04L_{\star,\mathrm{UV}}^{z=3}$ 
and integrate to the same numerical lower limit at all redshifts. The lower
limit is, of course, somewhat arbitrary but is designed to find a reasonable
medium between including a large fraction of the total luminosity/SFR density,
and preventing (possible over-) extrapolation by integrating to zero. In this
sense, it reflects the observational limits of the UV surveys. 

Admitting that this number is somewhat arbitrary, we adopt the same approach 
and use $0.04L_\star^{z=3} - \infty$ as the range for all of integrations of 
the UV LF. For $M_\star^{z=3} = -21.0$ (AB), the corresponding lower
luminosity limit is $4.36 \times 10^{27}$\ergsechz\ (unobscured SFR=0.6\msunyr).
By adopting this limit, our results can easily be cross-checked against the
available literature. At redshift 3 for the UV LF of \citet{Reddy2008}, this
range incorporates 70\% of an infinite integration under the LF.

Deciding upon a lower limit for the \halpha\ LF is more tricky, since it is
difficult to know if we are extracting comparable samples of galaxies. There is
no available $z=3$ \halpha\ LF, but if we adopt that compiled at $z=2.2$ in
\citet{Hayes2010a}, and set the lower limit to $0.04 L_\star$, we obtain a
luminosity of $4.6\times 10^{41}$~\ergsec. This corresponds to much higher
unobscured SFR than the lower UV limit at 3.5\msunyr. However, the UV and
\halpha-selection functions naturally recover galaxies of different dust
contents; if we translate these limits to ``true" SFRs for the respective
samples, we obtain limits of 2.6 and 6.0\msunyr\ for the UV and \halpha,
respectively. These limits differ by a factor of over 2 in SFR, but still are
not able to account for the differing populations of galaxies that survive the
respective selection functions -- were the dustier galaxies that are selected 
by \halpha\ able
to enter the UV-selected catalogues, the increased average dust content would
bring these values even closer together. We also argue that to some extent,
the overall shape of the UV and \halpha\ LFs must be governed by the same
physical processes and, regardless of the exact dust content, selecting 
galaxies brighter
than a certain fraction of the characteristic luminosity should recover objects
with similar underlying SFRs. Ultimately this argument is backed up in 
\S~\ref{sect:comp} when we find very similar UV- and \halpha-derived SFRs in the
local universe, and by the very similar SFR densities derived by the two tracers
in \citet{Reddy2008} and \citet{Hayes2010a}. Naturally by cutting both LFs at
the same fraction of $L_\star$, we recover similar fractions of the
luminosity density compared with integration to zero (70~\%). 

For \lya, the situation is more complicated still: cutting at the same intrinsic
SFR would mean that we do not include \lya\ emission at lower luminosities. This
is now not simply a matter of dust attenuation but also includes radiation transport
effects. Since we expect the line to be systematically weakened, applying a cut
at the corresponding SFR to that of \halpha\ or the UV would cause us to miss
much of 
this light. The best way to proceed, therefore, is to adopt the same philosophy
as above, and adopt $0.04L_{\star,{\mathrm{Ly}\alpha}}^{z=3}$. By selecting the
LF of \citet{Gronwall2007}, we obtain a lower limit of 
$1.75\times 10^{41}$~\ergsec. Should \fesclya=1, this would correspond to an SFR
of just 0.15\msunyr. However, in \citet{Hayes2010b} we determined a volumetric
\fesclya\ of just 5~\%, and scaling this SFR up by a factor of 20 brings it to
2.9\msunyr, almost perfectly into line with the UV-derived 2.6\msunyr\ discussed
above. Naturally, this integration from $0.04L_\star$ again includes 
$\approx 70$\% of the total luminosity density (compared with integrating from
zero). 

In summary, selecting the optimal integration limits is a non-trivial process,
yet we argue that by adopting these limits we should be selecting very similar
samples of galaxies, at least with respect to their unobscured SFR. The lower
limits are $4.36 \times 10^{27}$\ergsechz\ (UV), $4.6\times 10^{41}$~\ergsec\
(\halpha), and $1.75\times 10^{41}$~\ergsec\ (\lya). 
We have insured that these limits include the bulk 
of the luminosity density but are not dominated in uncertainty by extrapolation 
in the faint end, although we have also confirmed that integration to zero in
fact has only very minor effects on the final measurements of \fesclya.

\subsection{Compilation of the samples}\label{sect:comp}

All of the assembled data and the derived \fesclya\ measurements are summarized in 
Table~\ref{tab:fescz} and Figure~\ref{fig:fescz}. The measurements of \ebv\
relevant to each of the \halpha\ or UV measurements are derived from data in the
same publication as the \halpha\ or UV LF data themselves (with one exception, which
is discussed in the following paragraph).  In this subsection we 
provide the necessary motivation for our choices and comments on the various 
samples.

No instrumentation can perform a \lya-selected survey in the very nearby 
universe so we begin at $z\approx 0.2-0.4$ with the \lya\ LFs presented in both 
\citet{Deharveng2008} and \cite{Cowie2010}.
At these redshifts \halpha\ LFs are available, and therefore we proceed using 
Equation~\ref{eq:fesc_def_ha}. 
We adopt the \halpha\ LF of \citet{TresseMaddox1998}, and correct it for dust
attenuation by applying the 1~magnitude of extinction that is representative 
of local \halpha-selected galaxies \citep{Kennicutt1992}. 
For security and consistency with higher redshift measurements, we also examine
the $z=0.3$ UV LFs of \citet{Arnouts2005}, which we correct for dust using the
method of \citet{Meurer1999} and the $\beta$ slope measured by 
\citet{Schiminovich2005} in the same sample, finding extremely consistent
numbers.

Beyond the very nearby universe, no further \lya\ information is available before $z=2$, 
where we adopt our own measurement of 
\fesclya$=5.3\pm3.8$~\% \citep{Hayes2010b}, based upon \halpha\ and individually
estimated \ebv. 

It is already at this juncture in redshift that we lose the possibility to 
use \halpha, and therefore we proceed using published UV LFs and 
Equation~\ref{eq:fesc_def_uv}. Our next step is to take the \lya\
LF of \citet[][ $\langle z \rangle=2.5$]{Cassata2010} which we contrast 
against the dust-corrected \rholuv\ of 
\citet[][ $\langle z \rangle=2.3$]{Reddy2008}. 
For this, and all subsequent points from \citet{Cassata2010}, we adopt the 
values of $L_\star$ that are uncorrected for IGM attenuation. 
It is reassuring that the measurements at $z=2.2$ and $z \approx 2.5$ (which 
are based upon \halpha\ and UV, respectively) give very consistent numbers. 
Furthermore, in a very recent submission \citep{Blanc2010} an additional
\lya\ LF has been presented at $1.9<z<2.8$, the integrated \lya\ luminosity
density from which differs from our own result by $\approx 25$~\%. 

We then continue with the \citet{Reddy2008} UV data at $\langle
z\rangle=3.05$, which we use to compute \fesclya\ for the  $z=3.1$ \lya\ 
samples of \citet{Gronwall2007} and \citet{Ouchi2008}.

At $z\sim 4$ we have available \lya\ LFs from 
\citet[][ $z=3.7$]{Ouchi2008} and \citet[][ $z=3.9$]{Cassata2010}, and UV LFs
from \citet[][ $\langle z\rangle =3.8$]{Bouwens2007}. 
We also use the $z=4.5$ and 4.86 \lya\ LF points from \citet{Dawson2007} and
\citet{Shioya2009}, which we normalize by the dust-corrected UV point at 
$z=4.7$ from \citet{Ouchi2004}. 

The next redshift to examine is the popular $z\approx 5.7$ \lya\ window. Here 
we adopt the UV datapoint from the $i-$dropout sample of 
\citet[][ $\langle z\rangle =5.9$]{Bouwens2007}, and the 
\lya\ LF \citet[][ $\langle z\rangle=5.7$]{Ouchi2008}, which is in
good agreement with those of \citet{Shimasaku2006}, \citet{Ajiki2006}, and
\citet{Tapken2006}. We also add the highest redshift LF from 
\citet{Cassata2010} at $\langle z\rangle=5.65$.

Finally we assemble a few $z>6$ samples.  We adopt the $z=6.5$ point from
\citet[][which includes the sample of \citealt{Kashikawa2006}]{Ouchi2010}, 
and the measurement of \citet{Iye2006} at $z=7.0$, which
has also been compiled in \citet{Ota2008}. 
Here we adopt \cite{Bouwens2010} UV measurement at $\langle z \rangle=6.8$ for 
comparison. It should be noted that at this redshift the
dust-corrected and uncorrected measurements of \citet{Bouwens2010} 
converge. 
We adopt the most optimistic estimate at $z=7.7$ from \citet{Hibon2010}, 
for which we interpolate between the \citet{Bouwens2010} $z=6.8$ and 8.2 UV 
datapoints. \citet{Hibon2010} present Schechter parameters for four \lya\ 
LFs, based upon various assumptions about the rate of contamination by lower
redshift galaxies. By assuming all of their candidates are real (their sample 
$a$)  we find a \lya\ escape fraction of $(33.5^{+50.6}_{-33.5})$~\%. 
We also briefly 
examine their subsample $b$, in which only four of the seven objects are real.
For all of their sub-samples, the numbers are insufficient to provide 
meaningful errors on the luminosity density and by our standard error 
procedure we derive \fesclya=$(22.2^{+ 1707}_{-22.2})$~\%. Further, it should be noted
that in the \citet{Hibon2010} sample, the lower limits obtained on the \lya\ 
equivalent width are in the range 6--15\AA, with their continuum-detected object
showing \wlya=13\AA. Thus, at an acceptable confidence limit, none of their 
seven objects would actually survive the canonical \wlya\ cut of 20\AA\ that is 
typically employed in narrowband surveys. Including these data is therefore not 
straightforward, but in order to treat them as consistently as possible with
the lower redshift points, we have to set the $z\approx 7.7$ \lya\ escape 
fraction to zero, but adopted a characteristic error of 50.6~\% as derived from 
their most optimistic sample. We note that this limit is likely extremely high.

All of our measurements of \fesclya\ are listed in Table~\ref{tab:fescz} and 
shown graphically in Figure~\ref{fig:fescz}, which is the main result of this
paper.

\begin{deluxetable*}{llr|lllr|rl}
\tabletypesize{\scriptsize}
\tablecaption{Lyman-alpha escape fractions with redshift}
\tablewidth{0pt}
\tablehead{
		\multicolumn{3}{c|}{\lya\ quantities} & \multicolumn{4}{c|}{Intrinsic quantities} & \multicolumn{2}{c}{Derived results}\\
    \colhead{$z$}  & 
		\colhead{Ref} & 
		\colhead{\rhodstar } & 
		\colhead{$z$} & 
		\colhead{Ref} & 
	 	\colhead{\ebv } & 
		\colhead{\rhodstar } & 
	 	\colhead{\fesclya [ \% ]} & 
		\colhead{Comment} \\
    \colhead{(1)} & 
		\colhead{(2)} & 
		\colhead{(3)} & 
		\colhead{(4)} & 
		\colhead{(5)} & 
	 	\colhead{(6)} & 
		\colhead{(7)} & 
	 	\colhead{(8)} & 
		\colhead{(9)}
}
\startdata
\multicolumn{7}{l}{Estimates based upon \lya\ and \halpha\ luminosity functions ..........} \\
0.2--0.35 & De~08 & $(3.79\pm 1.69)\times10^{-4}$ & 0.2--0.35            & TM~98 & 0.33 & $(0.0303\pm 0.017)$ & $(1.25 \pm 0.90)$     & 1~mag at \halpha\\
0.2--0.4  & Co~10 & $(8.33\pm 2.60)\times10^{-5}$ & 0.2--0.35            & TM~98 & 0.33 & $(0.0303\pm 0.017)$ & $(0.275\pm 0.18)$     & 1~mag at \halpha\\
2.2       & Ha~10 & \nodata                       & 2.2                  & Ha~10 & 0.22 & \nodata             & $(5.3  \pm 3.8 )$     & Multi dimensional M.C.\\
\\
\multicolumn{7}{l}{Estimates based upon \lya\ and UV luminosity functions ..........} \\
2.5       & Ca~10     & $(7.08\pm 0.81)\times10^{-3}$ & $\langle2.3 \rangle$ & Re~08 & 0.15  & $(0.201\pm 0.022)$  & $(3.51\pm 0.56)$       &  \\
3.1       & Gr~07     & $(8.50\pm 5.32)\times10^{-3}$ & $\langle3.05\rangle$ & Re~08 & 0.14  & $(0.116\pm 0.017)$  & $(7.33\pm 4.71)$        &  \\  
3.1       & Ou~08     & $(5.54\pm 2.91)\times10^{-3}$ & $\langle3.05\rangle$ & Re~08 & 0.14  & $(0.116\pm 0.017)$  & $(4.78\pm 2.61)$        &  \\ 
3.7       & Ou~08     & $(4.78\pm 1.14)\times10^{-3}$ & $\langle3.8 \rangle$ & Bo~09 & 0.14  & $(0.089\pm 0.011)$  & $(5.36\pm 1.43)$        &  \\
3.8       & Ca~10     & $(8.71\pm 1.00)\times10^{-3}$ & $\langle3.8 \rangle$ & Bo~09 & 0.14  & $(0.089\pm 0.011)$  & $(9.77\pm 1.64)$        &  \\
4.5       & Da~07     & $(3.22\pm 1.25)\times10^{-3}$ & $\langle4.7 \rangle$ & Ou~04 & 0.075 & $(0.025\pm 0.011)$  & $(12.6\pm 7.17)$        &  \\ 
4.86      & Sh~09     & $(2.35\pm 3.17)\times10^{-3}$ & $\langle4.7 \rangle$ & Ou~04 & 0.075 & $(0.025\pm 0.011)$  & $(9.24\pm 13.0)$        &  \\ 
5.65      & Ca~10     & $(8.53\pm 3.44)\times10^{-3}$ & $\langle5.9 \rangle$ & Bo~09 & 0.029 & $(0.022\pm 0.005)$  & $(38.1\pm 17.2)$        &  \\
5.7       & Ou~08     & $(6.76\pm 4.77)\times10^{-3}$ & $\langle5.9 \rangle$ & Bo~09 & 0.029 & $(0.022\pm 0.005)$  & $(30.2\pm 22.2)$        &  \\
6.6       & Ou~10     & $(4.73\pm 1.24)\times10^{-3}$ &         6.5          & Bo~07 & 0.012 & $(0.016\pm 0.008)$  & $(30.0\pm 17.8)$        & UV Interpolated \\
7.0       & Iy~06     & $(1.07\pm 1.16)\times10^{-3}$ &         7.0          & Bo~09 & 0.010 & $(0.012\pm 0.008)$  & $(8.96\pm 11.5)$        & UV Interpolated \\
7.7       & Hi~10 & $(0 ^{+88.5}_{-0})\times10^{-3}$ &         7.7          & Bo~10 & 0.0   & $(0.005\pm 0.002)$  & $(0 ^{+50.6}_{-0})$         & UV Interpolated \\

\enddata
\tablecomments{For the \halpha-based estimates, we use the integrated luminosity
densities directly; SFRD measurements are presented just for homogeneity with 
the UV estimates. \rhodstar\ units of are \msunyrvol\ and \ebv\ is in magnitudes. 
The references are expanded as: 
Bo~09=\citet{Bouwens2009a};
Ca~10=\citet{Cassata2010};
Co~10=\citet{Cowie2010};
Da~07=\citet{Dawson2007};
De~08=\citet{Deharveng2008};
Gr~07=\citet{Gronwall2007};
Ha~10=\citet{Hayes2010b};
Hi~09=\citet{Hibon2010};
Iy~08=\citet{Iye2006};
Ou~04=\citet{Ouchi2004};
Ou~08=\citet{Ouchi2008};
Ou~10=\citet{Ouchi2010}; 
Sh~09=\citet{Shioya2009};
Re~08=\citet{Reddy2008};
TM~98=\citet{TresseMaddox1998}.
References for \ebv\ measurements are the same as for the
intrinsic star-formation rate density (I.e. that listed in the 5th column) with
the exception of the $\langle z\rangle = 0.3$ points in which \ebv\ is adopted
from \citet{Kennicutt1992}.  }
\label{tab:fescz}
\end{deluxetable*}

\subsection{Consistency (and inconsistency) between groups}\label{sect:consistency}

It should always be borne in mind that we are compiling results from 
different survey teams, who may adopt different techniques for data reduction
and photometry, derivation of the luminosity functions, and incompleteness
corrections. For example, \citet{MalhotraRhoads2004} find reasonable agreement
at $z\approx 5.7$ between the narrowband-selected \lya\ LFs of 
\citet{RhoadsMalhotra2001,Ajiki2004} and the lensing-based survey of 
\citet{Santos2004a}. However, the $z=5.7$ LF of \citet{Shimasaku2006}, on which
the study of \citet{Kashikawa2006} is based (see \S~\ref{sect:discdown}), 
find a strong disagreement at the
faint end between their own LF and the compilation of 
\citet{MalhotraRhoads2004}. As commented by \citet{Shimasaku2006} the likely 
cause for this discrepancy lies in (a) the lack of incompleteness corrections,
which are unmentioned in any of the 2004 articles, (b) the differences in 
equivalent-width based selection criteria, and (c) the large cosmic variance 
which \citet{Ouchi2008} noted can be of factors of $\approx 2$ in fields as
large as 1 square degree. Our results are sensitive to all of these 
considerations. 

It is only now that sufficiently large samples of \lya-emitting galaxies are 
presented in the literature for this study to be undertaken, and we 
are now fortunate that a good fraction of our data must contain internal 
self-consistency. For 
example, three of our data-points at (at $z=3.1$, 3.7, 5.7) are drawn from a 
single paper \citep{Ouchi2008} in which the methodologies must be internally 
consistent, and the basic trend can be seen in these data alone. A fourth point
at $z=6.6$ comes from \citet{Ouchi2010} where similar self-consistency is to be 
expected. In the same fashion, three further points are taken from 
\citet{Cassata2010} where internally the same methodology must have been 
adopted at each redshift. It is certainly encouraging that, for example at 
$z=5.7$ the measurements of \fesclya\ based upon \citet{Cassata2010} and 
\citet{Ouchi2008} are practically indistinguishable, despite the fact that 
they are based upon completely different methods: blind spectroscopy and 
narrowband imaging, respectively. The $z=2.2$ and 2.5 points of 
\citet{Cassata2010} and \citet{Hayes2010b} are similarly indistinguishable, as
(and also robust against the same fundamental methodological 
difference of blind spectroscopy vs narrowband imaging), are the $z=3.1$ points of 
\cite{Ouchi2008} and \citet{Gronwall2007} (both narrowband imaging). 

Any study of the galaxy population benefits by targeting spatially 
disconnected, independent pointings in order to beat down cosmic variance. By 
adopting the studies of various authors pointed all over the 
extra-galactic sky, this study is able to benefit from the inclusion of a 
large number of independent fields.

\vspace{5mm}

\section{General results} \label{sect:genres}

\subsection{The evolution of \fesclya }

\begin{figure*}[t!]
\includegraphics[angle=0,scale=1]{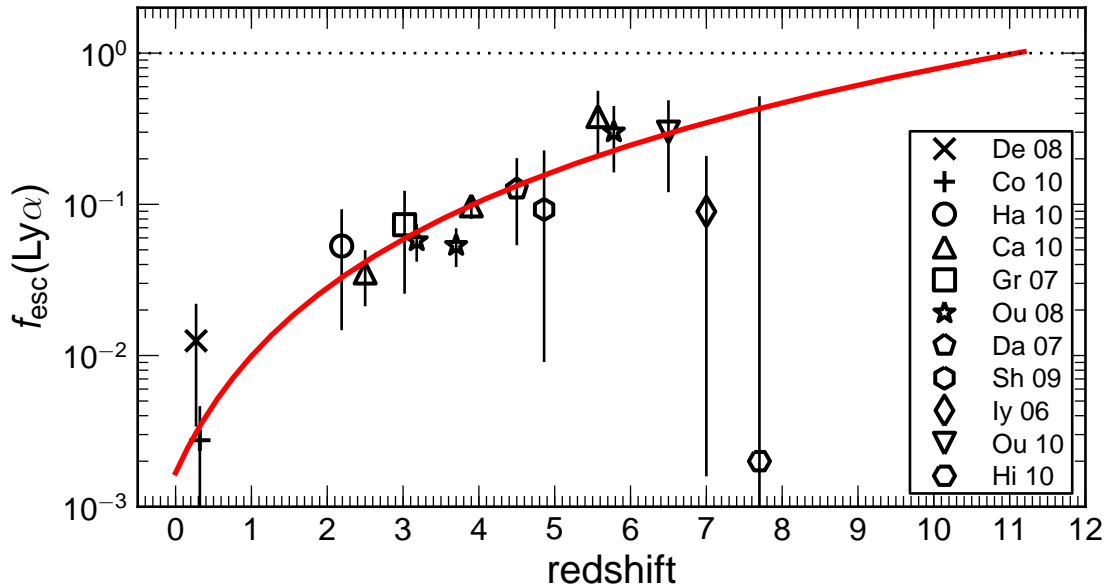}
\caption{The redshift evolution of \fesclya. Publication codes are listed in 
the footnote to Table~\ref{tab:fescz}.
$z=3.1$ and 5.7 points have been artificially shifted by $\Delta z=0.08$ 
for clarity. The point from \citep{Hibon2010} takes, according to our 
definition, a value of zero. It is therefore displayed at a value of 0.002 to permit
visualization on a logged axis.
The solid red line shows the best fitting power-law to points 
between redshift 0 and 6, which takes an index of $\xi=2.6$ and is clearly a
good representation of the observed points over this redshift range. It
intersects with the \fesclya=1 line (dotted) at redshift 11.1.  }
\label{fig:fescz}
\end{figure*}

Figure~\ref{fig:fescz} reveals a general and significant trend for \fesclya\ to
increase with increasing redshift. Beginning in the very local universe we see
\fesclya$\sim 0.01$ or lower for nearby star-forming objects. This increases to around 
$\approx 5-10$~\% by redshift of $\approx 3-4$, and further to 
$\approx 30-40$~\% by redshift 6. In order to quantify this trend we fit an
analytical function to these data-points, choosing a power-law of the form 
\fesclya$(z) = C\cdot(1+z)^{\xi}$ -- we obtain 
coefficients of $C=(1.67_{-0.24}^{+0.53})\times 10^{-3}; \xi= (2.57_{-0.12}^{+0.19})$. 
Note that we do not include any $z>6$ points in our fit since previous studies
suggest that it is around this redshift that an appreciable fraction of the
intergalactic hydrogen becomes neutral, and may in principle affect the \lya\
LF. For more discussion on this see \S~\ref{sect:discdown}.
To insure that the fit is not biased by the presence of two $z\approx 0.3$ 
points that lie around 8~Gyr from $z\approx 2$, we repeat the fit 
after excluding these points, finding 
$C=(4.79 _{-0.69}^{+5.68})\times 10^{-4}; \xi= (3.38 _{-0.37}^{+0.10})$. 
Clearly the fit is affected by these points, but their exclusion actually 
results in a more rapid evolution with redshift. 

Beyond redshift 6 the apparent trend begins to break but it is initially very
slow. Over the redshift interval of 5.7 to 6.5, \fesclya\ stabilizes, but
decreases again to just $\approx 10$~\% at $z=7$. The redshift 7 point 
from \citet{Iye2006} is confirmed, whereas none of the sample 
of redshift 7.7 candidates from \citet{Hibon2010} have confirmations by
spectroscopy, and this upper errorbar must be regarded as an
optimistic upper limit.

Finally, we perform a simple experiment with the best-fit relationship to the 
\fesclya$-z$ trend, and extrapolate to estimate the redshift at which \fesclya\
reaches unity. This would carry the implication that the ISM of the average
galaxy has become effectively devoid of dust, and since dust is a byproduct of
the star-formation process, must also correspond to a time of approximately
primeval star formation. It is interesting, therefore, that we find \fesclya=1 
at $z=11.1_{-0.6}^{+0.8}$, which is consistent 
with the redshift of the instantaneous reionization of the Universe based upon
{\em W-MAP} data \citep[$z=11\pm1.4$;][]{Dunkley2009}.

\subsection{Comparison with the literature}\label{sect:fesc_comp}

Naturally this is not the first time that \fesclya\ has been estimated and 
several other studies based on a wide array of methods have attempted to 
pin down the same quantity at different redshifts. 

For example, at redshifts of 5.7 and 6.5, we compute \fesclya\ of around 40~\%
and 30~\%, respectively. Based upon the fitting of spectral energy distributions
(SED) to stacked broadband fluxes, \citet{Ono2010} estimate 
\fesclya$=(36^{+68}_{-35})$~\% and $(4 ^{+180}_{-3.8})$ at the same redshifts. 
Although derived from an interesting approach, the uncertainties
are still too large to provide a useful comparison.

Like us, \citet{Nagamine2008} compared observed \lya\ LFs 
\citep[][in this case, which we also use]{Ouchi2008} with intrinsic 
estimates, having derived this intrinsic LF from smoothed particle 
hydrodynamical (SPH)  models of galaxy formation. They adopt two methods of scaling
the intrinsic to the observed LFs, the first of which they call `escape 
fraction', which is a scaling to the datapoints along the luminosity axis, 
and assumes all galaxies have the same \fesclya. This method finds 
\fesclya=10~\% at $z=3$, which is certainly consistent with our estimates 
based on the $z=3.1$ LF of \citet{Gronwall2007} and similar to, but slightly 
higher than our estimate based on \citet{Ouchi2008}. At $z=6$ however, 
\citet{Nagamine2008} require an escape fraction of just 15~\% which is 
lower than our estimates of $30-40$~\%, and discrepant with our estimates
at around the $2\sigma$ level. 
\citet{Nagamine2008} also test a 
`duty-cycle' scenario (an LF scaling along the $\Phi$ axis) 
in which only a fraction of the SPH galaxies are `on'
as \lya-emitters but emit 100~\% of their \lya\ photons. Note that in these two
extreme scenarios, there is no requirement for the integral over the scaled LF to be
equivalent. \citet{Nagamine2008}
present duty cycles of 0.07 and 0.2 at $z=3$ and 6, respectively.
However, before they compute these scalings the observed LFs are shifted along
the luminosity axis by IGM attenuation factors of 0.82 ($z=3$) and 0.52 ($z=6$), 
which also need to be applied for a comparison with our estimate.  
Thus in the duty
cycle scenario, the volumetric escape fractions that one would infer from 
the study of \citet{Nagamine2008} are 6~\% at $z=3$ and 10~\% at $z=6$. Again
this agrees very well with our measurement at $z\approx 3$ but 
compared with our estimates at $z=6$ is an underestimate of around the 
same magnitude as their escape fraction method. 

In contrast, using similar SPH galaxy formation models but modified 
prescriptions for \lya\ production and transmission, as well as a different
reionization history, \citet{Dayal2009} find \lya\ escape fractions of 
30~\% at both $z=5.7$ and 6.5, which corresponds exactly with our 
measurements. 
Similar values of \fesclya\ $\sim$ 23--33\% have also been obtained 
in the follow-up work of \citet{Dayal2010}, although they include also
an IGM transmission of $T_\alpha=0.48$. Throughout this paper we have 
made sure not to apply any IGM correction, since the value of $T_\alpha$ 
remains 
poorly constrained, even theoretically, and from an observational perspective 
there is no strong evidence for exactly how close the IGM comes to a narrow 
\lya\ line. As with the \citet{Madau1995} prescription, it is likely that this 
IGM transmission is too low when considering lines that are systematically 
redshifted by 
the kinematics of the ISM, which would drive up these theoretical estimates of 
the \lya\ escape fraction. 

Adopting a similar method of LF scaling by luminosity, \citet{leDelliou2005}
found that an escape fraction of 2~\% was sufficient to match observed \lya\ 
LFs with their predictions based upon semi-analytical 
models between $z=2$ and 6, with the same machinery able to predict the
clustering properties of \lya\ emitters \citep{Orsi2008}. 
This is at the lower end of being consistent with 
our $z=3$ measurements, and should the same escape fraction hold at $z=0.3$, 
would also be consistent with our estimates in the nearby universe. 
However, the \citet{leDelliou2005} escape fraction is highly inconsistent
with our estimates at higher redshift. These semi-analytical models, using the 
prescription
of \citet{Baugh2005}, categorized star-formation as occurring in two discrete 
modes, with a normal Salpeter IMF $(\alpha=-1.35)$ assigned to quiescent 
star-formation and a flat IMF $(\alpha=0)$ for bursting systems. This flat IMF
increases the ionizing photon production at a given SFR by a factor of ten
and was implemented as a requirement in order to reproduce the population of 
sub-mm selected galaxies at $z>2$. 
However as noted by \citet{leDelliou2006}, the fraction of total star-formation
that occurs in bursts increases from 5~\% at $z=0$ to over 80~\% at $z=6$, and
thus their model implies that by the $z=5.7$ points, effectively all stars
are formed in environments where ionizing photons are greatly over-produced 
compared
to the present day. However, should this requirement of the flat IMF be removed
and Salpeter applied throughout,
the intrinsic rate of production of ionizing photons would be decreased by a 
factor of 3 at $z=3.1$ where the star-formation is shared evenly between 
bursting and quiescent systems. This would bring the \fesclya\ estimate
to 11~\% at this redshift. At $z=6$, \fesclya=16~\%  would be found by 
replacing the flat IMF with Salpeter. These numbers are indeed very similar
to the SPH models of \citet{Nagamine2008} but inconsistent with those of 
\citet{Dayal2009} and our own estimates based upon observation. It is 
interesting to point out, however, that the IMF assumption has little effect on
the $z\approx 0.3$ points where, in their model, the quiescent mode of 
star-formation dominates.

\subsection{Possible physical explanations}\label{sect:fesczbrief}

The evolution in measured \fesclya\ is substantial, covering approximately two
orders of magnitude, and no doubt holds vital information about the physical
nature of galaxies at various cosmic epochs. As we will show in
\S~\ref{sect:disc}, the most likely explanation for this evolution is the
decrease of the average dust content of galaxies.
However from a physical perspective many effects may enter. 
For example, galaxies may also contain less neutral hydrogen to scatter 
photons, show faster outflows, or become more clumpy. 
The inferred increase may alternatively be mimicked by galaxies becoming 
younger on average, having low and decreasing metallicities, or forming stars 
with IMFs that become more biased in favor of massive, ionizing stars. 
On the other hand, the scattering of \lya\ photons by a neutral IGM, and the
general leakage of ionizing photons (LyC) are expected to increase with
increasing redshift, and would both serve to lower the perceived \lya\ escape fraction 
(although the ``true'' \fesclya\ of galaxies, i.e.\ before the IGM, would not be affected).
		
Regrettably we are not able to measure any of 
these quantities directly from this compilation of data. 
We have, however, assembled data that show a number of trends with redshift: 
the \lya\ and UV luminosity densities and the dust contents. These we have 
combined to show how \fesclya\ evolves, yet in order to 
extract the maximum of information from these, we need to examine another
possible trend: how \fesclya\ correlates with dust content. Thus we delay
a detailed discussion of what drives the \fesclya--$z$ trend until 
\S~\ref{sect:disc} and now proceed to discuss the effects of radiation
transport and dust absorption. 

\vspace{5mm}

\section{The \lya\ escape fraction and its dependencies}\label{sect:fescebv}

That \lya\ photons undergo a complex radiation transport, in which a large 
number of parameters enter, is well-known but poorly understood from an 
empirical angle. 
Transport is thought to be affected by dust content \citep{Atek2008,Atek2009,Hayes2010b}, 
dust geometry \citep{Scarlata2009}, 
\hi\ content and kinematics \citep{Kunth1998,MasHesse2003,Shapley2003,Tapken2007}, and 
geometry/neutral--ionized gas topology 
\citep{Neufeld1991,Giavalisco1996,HansenOh2006,Finkelstein2008,Finkelstein2009}.
Unfortunately, \hi\ masses remain impossible to measure directly beyond
the very local universe. Kinematic measurements of the neutral ISM can be obtained at 
high-redshift, but require deep absorption line spectroscopy against the 
vanishing continuum of \lya-selected galaxies and thus are prohibitively 
expensive for large samples of individual galaxies.
We are therefore effectively limited, when 
targeting statistically meaningful samples, to examining \lya\ emission against 
the dust content, and have to infer information about the remaining quantities 
by secondary analysis. 

\begin{figure}[t!]
\includegraphics[angle=0,scale=0.55]{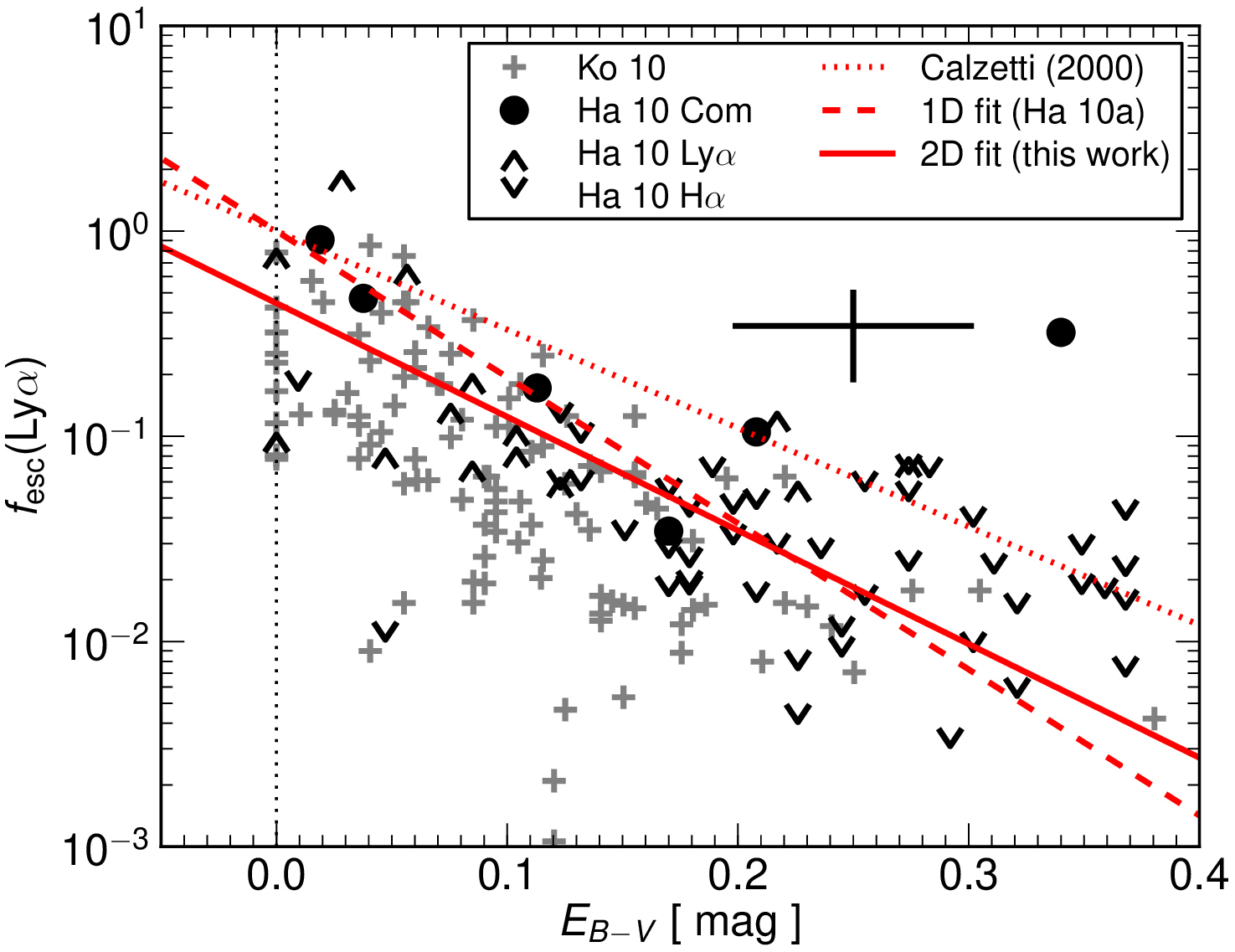}
\caption{Literature compilation of \fesclya\ vs \ebv. The codings in the legend
are: 
Ha~10=\citet{Hayes2010b};
Ko~10=\citet{Kornei2010}.
Solid circles from \citet{Hayes2010b} are six objects for which we have 
detections in both \lya\ and \halpha. Caret down markers are \halpha\ emitters that
were undetected in \lya\ and hence presented as upper limits, while caret up
markers are \lya\ galaxies for which \halpha\ lies below the detection limit and are 
hence presented as lower limits. Errorbars are removed from the plot to aid 
readability, but the average errors from the common detections of \citet{Hayes2010b} 
are shown by the singular black point with errorbars. For further information the 
reader is referred to Figure~3 of \citet{Hayes2010b}. 
The red lines show various conversions between the observed stellar \ebv\ and
\fesclya. The dotted line shows the standard \citet{Calzetti2000} prescription,
the dashed line shows the 1 dimensional fit to the data from \citet{Hayes2010b}
and the solid line a 2 dimensional fit described in the text. 
\medskip
}
\label{fig:fescebv}
\end{figure}

Significant anti-correlations between \fesclya\ and \ebv\ have been presented 
in four recent papers, all of which invoke different selection functions and 
employ
different methods of analysis. Firstly, \citet{Verhamme2008} used radiation 
transport modelling of spectrally resolved \lya\ features in a sample of LBGs 
between redshift 2.8 and 5 to estimate both dust attenuation and \fesclya. Based
upon the Balmer line ratio (\halpha/\hbeta), \citet{Atek2009} computed \fesclya\
and nebular reddenings based upon purely nebular physics in a sample of nearby 
\lya-selected galaxies. Were \halpha\ and \hbeta\ observations available in the 
distant universe, this method would be the ideal one by which to proceed. More
recently, \citet{Kornei2010} performed a similar experiment in a sample of
redshift $\sim 3$ \lya-emitting LBGs, in which dust attenuation and intrinsic
\lya\ luminosities were estimated from modelling of the SED. Finally in sample of redshift 2
\lya- and \halpha-selected galaxies, we also used SED modeling to estimate \ebv\
but estimated the intrinsic  \lya\ production from the dust-corrected \halpha\ luminosity
\citep{Hayes2010b}.

In Figure~\ref{fig:fescebv} we show a compilation of the \fesclya\ and \ebv\
points from \citet{Kornei2010} and \citet{Hayes2010b}. Here we adopt only these
two data-sets since they involve similar computations of \ebv\ but include \lya,
\halpha, and UV selection and should be broadly representative of
the general galaxy populations under consideration in this paper. 
These two studies both perform full SED fits, but use them in different 
ways, with \citet{Kornei2010} requiring the intrinsic ionizing photon budget
to estimate \fesclya\ and \citet{Hayes2010b} using only the \ebv\ estimate to 
correct \halpha\ for the dust attenuation. Thus the \citet{Kornei2010} points
are in principle expected to be more sensitive to the standard set of 
assumptions in population synthesis (IMF, stellar atmosphere models, etc). 
However a substantial overlap between the two populations is 
clear in Figure~\ref{fig:fescebv}, with the two populations occupying a very 
similar region of the \fesclya--\ebv\ plane (the fact that we find more 
galaxies at higher \ebv\ is due to the fact we find redder galaxies by 
\halpha\ selection than is possible using the UV-biased Lyman-break criterion).

The dotted line shows the dust attenuation prescription of \citet{Calzetti2000}
which should be valid in the case of no \lya\ scattering and a simple dust
screen. This line is described by 
$f_\mathrm{esc}^{\mathrm{Ly}\alpha}=10^{-0.4\cdot E_{B-V}\cdot 
k_{1216}}$, where $k_{1216}=12$.
Very few points lie above this line and all are likely placed there by
statistical scatter. Indeed, this line sets an approximate upper limit to the
datapoints, which extends in the direction of lower \fesclya\ due to radiation
transport effects increasing the effective dust optical depth seen by \lya. 

In attempts to quantify the effects of resonance scattering and dust absorption, 
the studies of \citet{Verhamme2008}, \citet{Atek2009}, and \citet{Hayes2010b} all
fit linear relationships to the datapoints on the log(\fesclya)--\ebv\ plane, 
assuming no
a priori information about the dust. These studies all used a functional 
form of $f_\mathrm{esc}^{\mathrm{Ly}\alpha}=10^{-0.4\cdot E_{B-V}\cdot 
k_{\mathrm{Ly}\alpha}}$, where \klya\ (the single free parameter of the fit)
is an effective extinction co-efficient 
for \lya, and thus includes both scattering and absorption. Both at high-$z$,
the studies of \citet{Verhamme2008} and  \citet{Hayes2010b} found effectively
the same value of \klya=17.8, which runs significantly steeper than the
\citet{Calzetti2000} relationship as \lya\ photons are preferentially
attenuated. This is shown by the dashed line in Figure~\ref{fig:fescebv}. 

These formalisms force the fits to conform to \fesclya=1 at \ebv=0, and
technically it is true that if there is exactly zero dust, \lya\ photons cannot
be absorbed by dust. However, the very presence of \lya\ photons implies that 
star-formation must be occurring and, after just $\sim 3$~Myr of star-formation,
dust produced in supernovae would be returned to the ISM and the optical color
excess ceases to be a good proxy for dust. It is well-known that \lya\ can be strongly 
suppressed even when miniscule amounts of dust are present 
\citep[e.g.][]{Hartmann1984,Kunth1994,ThuanIzotov1997,Ostlin2009} and as 
Figure~\ref{fig:fescebv}
shows some galaxies have \fesclya=10\% with no measurable UV attenuation.
Indeed, many star-forming galaxies show little or no attenuation in front of
their ionizing clusters but substantially attenuated nebular regions. This
is the origin of the factor of 2.2 difference between stellar and nebular
measurements of \ebv\ \citep{Calzetti2000}, but at a very low UV stellar attenuation of 
\ebv$\approx 0$ applying a factor of two is not meaningful and nebular 
lines in general -- and \lya\ in particular -- may be heavily attenuated. 
It is unfortunate that at high-$z$ the UV continuum is our only proxy for the
dust content as we indeed expect to be surveying redshifts at which the stellar
attenuation indeed falls to $\sim 0$ \citep[e.g.][]{Bouwens2009a}.

To account for these factors we now proceed to relax the requirement of the fit passing 
through (\ebv,\fesclya)=(0,1) and re-fit the combined datasets of 
\citet{Kornei2010} and \citet{Hayes2010b} using the following expression 
\begin{equation}
f_\mathrm{esc}^{\mathrm{Ly}\alpha}=C_{\mathrm{Ly}\alpha} \cdot 10^{-0.4\cdot E_{B-V}\cdot
		k_{\mathrm{Ly}\alpha}}.
\label{eq:fescebv}
\end{equation}
This expression takes the same form as the standard dust-screen prescription,
with coefficient \klya, but adds the additional free parameter of \clya, the
factor by which \fesclya\ is scaled down. 
As in \citet{Hayes2010b} we use Schmidt's binned linear regression algorithm 
\citep{Isobe1986}, since it
permits the combination of data-points and limits in both directions. For \klya\
we obtain a value of 13.8, which is much more similar to the value
of 12.0 obtained from \citet{Calzetti2000} at the wavelength of \lya.
However, we also obtain \clya=0.445, indicating we expect \fesclya\ to be around
50~\%, even when there is no measurable dust attenuation on the stellar
continuum. This is in
fact a more plausible scenario since the effect of scattering by neutral 
hydrogen is not expected to depend on the dust content itself. 
This fit is shown by the solid red line in Figure~\ref{fig:fescebv}.
Again the points of \citet{Kornei2010} and \citet{Hayes2010b} are subject
to different assumptions that enter the population synthesis. However, for the
reasons outlined previously in this subsection and the similarity between the 
distributions, we do not expect these quantities to be strongly subject to these
assumptions.

It is not necessarily straightforward to define a goodness-of-fit measurement
to compare the quality of the three fits, given the large number of upper- and 
lower-limits in this dataset. Thus we define our own normalized r.m.s. statistic 
($rms_\mathrm{n}$), as: 
\begin{equation}
rms_\mathrm{n} = \sqrt{ \frac{1}{N} \sum _i ^N \left(\frac{f_i^\mathrm{meas} - f_i^\mathrm{EBV}}{f_i^\mathrm{meas}} \right)^2 } \,
\end{equation}
where  $f_i^\mathrm{meas}$ is the $i^\mathrm{th}$ measured \lya\ escape fraction, 
$f_i^\mathrm{EBV}$ is the $i^\mathrm{th}$ \lya\ escape fraction predicted from 
\ebv, and $N$ is the number of data-points. 
However in order to treat the limits, we permit a point to 
contribute to the summation only if that limit is violated. We appreciate that this 
is a non-standard statistic, but it does enable a quantitative measure of the 
goodness-of-fit that is philosophically not too far removed from more 
commonplace statistics. 
Adopting the \fesclya--\ebv\ relations derived from \citet{Calzetti2000}, the 
one parameter fit from \citet{Hayes2010b} and the two parameter fit from this 
work, we compute $rms_\mathrm{n}=1.85$, 1.02, and 0.66, respectively.

We have now assembled information about three trends: the observed redshift 
evolution of \fesclya; the observed redshift evolution of the dust content of 
galaxies; and the observed relationship between \fesclya\ and dust content.
We will next show that we are able to synthesize these points to infer some 
general trends in the evolution of galaxies.

\section{On the evolution of \fesclya}\label{sect:disc}

\subsection{Redshifts 0--6: the upwardly evolving escape fraction and the 
		properties of galaxies}\label{sect:discup}

\subsubsection{The evolving dust content of galaxies}\label{sect:discup:dust}

\begin{figure}[t!]
\includegraphics[angle=0,scale=0.55]{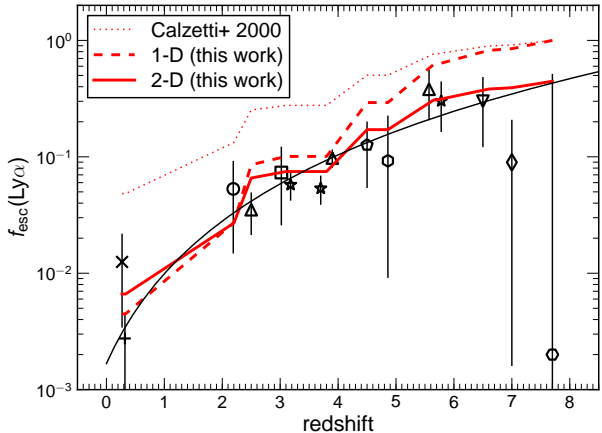}
\caption{Same as Figure~\ref{fig:fescz}, but zoomed onto the relevant region. 
The red lines show the \lya\ escape fractions that would be predicted based upon
the values of \ebv\ that have been measured in the respective \halpha\ and UV
samples (listed in Table~\ref{tab:fescz}), and using the various conversions 
between measured \ebv\ and \fesclya\ described in the text. The dotted line represents the dust
attenuation law of \citet{Calzetti2000}, the dashed line the 1 dimensional
empirical fit to the data of \citet{Hayes2010b}, and the solid line a 2
dimensional fit to the data described in \S~\ref{sect:fescebv}. Using the 2
dimensional fit, a remarkably good agreement is seen between observations and
prediction between redshifts 0 and 6.5.  }
\label{fig:fesczann}
\end{figure}

We showed in the previous section that \fesclya\ of individual galaxies is 
anti-correlated with the measured \ebv\ (Figure~\ref{fig:fescebv}). Given 
that the typical \ebv\ evolves with redshift (see Table \ref{tab:fescz}),
we may indeed expect a positive correlation between \fesclya\ and redshift.
This is exactly what Figure~\ref{fig:fescz} shows, where it is clear that the \lya\ 
escape fraction increases smoothly and monotonically out to $z\sim 6$. Thus it
appears that this increase in \fesclya\ is the result of the dust content of the
star-forming galaxy population decreasing with redshift.
We now take the measured values of \ebv\ from the various samples (listed in
Table~\ref{tab:fescz}), and use them to compute the \fesclya\ that would be
expected, from the three conversions between \ebv\ and \fesclya\ discussed in
the previous section [\citet{Calzetti2000}, an empirical fit with one free
parameter \citep{Hayes2010b}, and an empirical fit with two free parameters]. 
We show the measured escape fractions together with these
predictions in Figure~\ref{fig:fesczann}. 

We first discuss the predictions based upon the  
\citet[][red dotted line]{Calzetti2000},  which
is clearly discrepant with the observations at around the $3\sigma$ level at 
every redshift. Obviously this is to be expected since \lya\ photons resonantly
scatter and it is unlikely that the dust is distributed in a uniform screen. The
one dimensional fit from \citet{Hayes2010b} offers substantial improvement and
is able to describe the observations between redshifts 0 and 4. This
reasoning is circular for the redshift 2 points where the \fesclya--\ebv\
relationship was derived, but we stress the tautology is present only at this
redshift. This relationship is not able to explain any of the datapoints at
redshift above 4, where it systematically over-predicts the \lya\ escape
fraction. 

As redshift increases the dust content of galaxies is clearly shown to change
and, could we plot Figure~\ref{fig:fescebv} at redshifts higher than 3, we could
expect galaxies to cluster successively further towards the upper left corner of
the plot. Since the \citet{Hayes2010b} \fesclya--\ebv\ fit is forced through the
(\ebv,\fesclya)=(0,1) coordinate and a high value of \klya\ is found, the
predicted escape fraction evolves very quickly with redshift. Indeed, these 
predictions evolve much faster than the data, as \fesclya\ is forced for
unphysical reasons towards unity. 

When we introduce the new \fesclya--\ebv\ fit with two free parameters and
allow \clya$\ne 1$,
the agreement between the measured and observed \lya\ escape fractions is
striking: it agrees with essentially every datapoint, within the errorbars,
between redshift 0 and 6.6. We should point out that it is not clear that
the use of the average \ebv\ for a sample should by necessity reproduce the 
volumetric escape fraction. Due to variations of the dust contents and 
ISM of individual galaxies, and the associated impact upon the transfer of 
\lya\ and the selection of galaxies, it is possible that the average \fesclya\
could have been skewed substantially from the data-points. Indeed, close 
examination of the \fesclya--\ebv\ relationship (\clya=1; \klya=17.8) 
from \citet{Hayes2010b} reveals
that it does not perfectly intersect the center of the \fesclya\ datapoint 
($z=2.2$ point in Figure~\ref{fig:fesczann})
from the same survey, despite \fesclya, average \ebv, and the coefficients of 
the \fesclya--\ebv\ relationship all having been derived entirely from this one
dataset. This most likely results from the weighting across the population 
from which the average \ebv\ is computed
(the representative \ebv\ is not an average weighted by the intrinsic \lya\ 
luminosity), exactly the effect under discussion.  However, the fact 
that such tight agreement is seen between the observational estimates and those
derived from our fit suggests that such a bias in the selection of the 
populations is not at play here.

Again we stress
that the relationship we derived between \fesclya\ and \ebv\ in 
\S~\ref{sect:fescebv} includes the effects of resonance scattering, and thus 
in some manner the neutral gas content, its kinematics and relative geometry
all enter the relationship, which holds even when the measured optical color
excess on the stellar continuum is zero.
There is no reason to assume that these quantities
are constant with redshift and we could, for example, envisage situations where
the gas content, feedback properties, or clumpiness evolve and
thereby change \klya\ or \clya.
However the tight agreement between our observed \fesclya\ values and
those computed from the \fesclya-\ebv\ relationship provides no evidence for 
the evolution of these properties (at least if the gas content does change it does
not take part in the \lya\ scattering process). The evolution of \fesclya\ 
across almost the entire observable universe can be explained cleanly within 
the confines of this simple model, 
as mainly due to a dust content that evolves with redshift.

\subsubsection{Other effects}\label{sect:discup:other}

We need to interpret an increase in the global \fesclya\ of galaxies by a 
factor of $\sim 4$ between $z=2$ and 6, and naturally if something were to
alter the intrinsic \lya/UV ratio of galaxies by this factor, the evolution in 
\fesclya\ could be mimicked.

For example, there is evidence that the \wlya\ distribution of galaxies changes with 
increasing redshift: high-\wlya\ objects become relatively more abundant 
(e.g. \citealt{Gronwall2007} c.f. \citealt{Shimasaku2006}; also 
\citealt{Ouchi2008}), and thus pure selection may explain the trend. 
However, the \wlya\ distributions at $z=2$ and 3 suggest a maximum of 
$\sim 20$~\% of the total luminosity density will be lost by non-selection of 
$0<W_{\mathrm{Ly}\alpha}<20$~\AA\ galaxies, and such a selection bias can
certainly not explain the magnitude of the trend observed here.

It may also be argued that lower metallicities or a flattening of the IMF
may explain the trend. However, between solar and $1/50$ solar metallicity
the increase of \wlya\ for constant SFR, a measure of the relative \lya/UV 
output, is less than 50~\% \citep{Raiter2010}, insufficient to 
explain the observed increase of \fesclya.
To explain an increase by a factor $\sim 4$ would require a decrease
of the average metallicity from solar down to less than $10^{-3}$ solar
\citep{Raiter2010}, which seems highly unlikely. 

One would also assume that a relatively higher fraction of genuine primeval galaxies 
would be discovered as redshift increases, and a substantial ($\sim 3$-fold) 
enhancement of \lya/UV may arise from preferential selection of extremely young 
systems \citep[e.g.][]{CharlotFall1993,Schaerer2003}. To get this kind of
enhancement a galaxy must either be observed at an age below $\sim 10$~Myr or,
should an episode of star-formation occur superimposed atop an aged stellar
population, sufficient time must have elapsed for that population to fade in the
UV. For this UV fading to occur, punctuated bursts of star-formation would need 
to be separated by around the UV equilibrium timescale of $\sim 100$Myr.
At $z=6$ the Universe has an age of 1~Gyr and even if all star-formation
were to occur in individual bursts, the chance of catching an individual galaxy
at this time would be around 10~\%. Thus, integrated over the entire galaxy
population the application of such a sampling bias also seems quite implausible. 

We may expect at some point over this cosmic evolution, that galaxies
start to leak a substantial fraction of their ionizing photons (\fesclyc). 
Indeed as we
approach the middle of the epoch of reionization, the reionization processes 
itself dictates that this must be true, and we may expect at lower redshifts 
(e.g. 4--6) that a substantial population of galaxies may remain with an ISM 
that permits high \fesclyc. In addition, across approximately the same redshift 
domain we may
expect the thickening neutral phase of the IGM to start to suppress \lya. Both
of these effects would act to lower the perceived \lya\ escape fraction by
either draining ionizing photons or scattering \lya. Although we are not able to
tell whether these effects become significant at $z\sim 4-6$, if they do become
important then
the intrinsic \lya\ escape fractions of these galaxies will be still higher than we
measure\footnote{For example, assuming that half of the \lya\ flux is lost
due to scattering in the IGM the ``intrinsic'' value of \fesclya\ out of galaxies
would be higher by a factor 1.22 (1.92) at $z \sim$ 3 (6), assuming the 
average IGM opacity of \citet{Madau1995}.}.

It may be argued that the measured \lya\ fluxes (and hence the \lya\ 
luminosity density) could be underestimated due to the spatial extension of \lya, and that 
some of the observed redshift trend could be due to this 
\citep[e.g.][]{LoebRybicki1999,Zheng2010}. Although a somewhat 
larger spatial extension of \lya\ compared to the UV continuum has been noted 
in some surveys \citep[e.g.][]{Nilsson2009,Finkelstein2010}, stacking analysis 
in other Hubble Space Telescope images reveals the \lya\ emission to be 
spatially compact, with only a small fraction of the integrated luminosity lost 
to aperture effects \citep{Bond2010}. 
Therefore it seems very unlikely that this could lead to a significant 
underestimate of 
the \lya\ flux, which would mimic the apparent trend of increasing \lya\ 
escape fraction with redshift. The main reasons are the following.  First, the 
photometric apertures typically used for the narrowband images taken from the 
ground are several times larger than the FWHM of the \lya\ emission and several
studies apply the same method at several redshifts (e.g.\ between $z \sim 3$ 
and 6, \citealt{Ouchi2008}). Second, several independent measurements using both
imaging and spectroscopy reveal the same trend between $z \sim 2$ and 6 
\citep{Ouchi2008,Cassata2010,Stark2010a}, and also over a smaller redshift range
\citep{Reddy2008}. Third, it is well-known that in individual \lya-selected 
systems at redshifts 2--3, the SFR inferred by comparing \lya\ and UV radiation 
is frequently found to be comparable 
\citep{Guaita2010,Nilsson2009,Ouchi2008}.
Finally, some of the brightest \lya-emitting objects on the sky -- where the 
order-of-magnitude fainter low surface brightness scattered emission should 
become apparent -- also seem to be spatially 
compact \citep[e.g.][]{Westra2006}. These observational lines of evidence all
argue against an important loss of \lya\ photons related to its spatial extension.

At $z \sim 0.2-0.3$ the \lya\ emitting samples have been carefully constructed 
from surveys using GALEX slitless spectroscopy of NUV continuum selected objects
\citep{Deharveng2008,Cowie2010}. Given its relatively low spatial resolution 
($\sim 5\arcsec$) the \lya\ flux measurement of individual sources should not 
be affected by possible differences in the spatial extension. Furthermore, 
blending affects only 10\% of the sources, according to \citet{Cowie2010}.
Finally, comparing number counts of GALEX sources with/without \lya\ emission 
these authors have also shown that the \lya\ emitters represent only $\sim 5$~\%
of the NUV-selected continuum sources, a fraction significantly lower than the 
20--25\% derived for $z \sim$ 3 LBGs by \citet{Shapley2003}. In other words, a
low escape fraction at low-$z$ is not only obtained from the ratio of the UV and
\lya\ luminosity density, but also from direct inspection of NUV continuum 
selected objects.

Finally, as discussed in \S~\ref{sect:intlims}, our assumed limits of 
integration may introduce an overall bias into the data. For both the \lya-
and UV-selected populations, the characteristic luminosity of the LF 
($L_\star$) is known to evolve with redshift. Thus selecting a constant lower
limit at all redshifts may result in an artificial evolution. Firstly it 
should again be noted that our fixed lower limits apply to both the 
numerator and denominator (\lya\ and UV LFs; in Equation~\ref{eq:fesc_def})
and to first order will cancel. Secondly, the evolution of both \lya\ and UV 
LFs follows a similar pattern, starting low in the nearby universe and 
increasing rapidly to $z=2$ or 3, from where they begin to decline in the 
direction of the highest redshifts (with the \lya\ LF declining slower than that
of the UV in this range). Thus were this effect to be significant, and also not to 
cancel as just suggested, we would expect a strong upwards evolution from 
$z\approx 0$ to 2 which we do see, followed by a slow decline to higher 
redshift, which is certainly not reflected in the data.

In short, the various methods and arguments all point clearly towards a
significant evolution of the \lya\ space fraction with redshift. The main 
uncertainty affecting the precise absolute value of \fesclya\ is probably due 
to statistical uncertainties in the LFs and to the simple extinction 
correction applied to derive it, not possible \lya\ losses due to apertures.

\subsection{The downwardly evolving escape fraction and the properties of the
intergalactic medium}\label{sect:discdown}

Beyond a redshift of around 5.7, the measured value of \fesclya\ begins to 
decline, although 
initially this decline is weak and the deviation from our best-fit
relationship at redshift 6.5 is not significant. Adding the $z=6.5$ point
of \citet{Ouchi2010} and \citet{Kashikawa2006} to our fit does not change 
the result. 
However the $z=7$ point lies at just 8~\%, and is around $2\sigma$ below both
the best-fit \fesclya--$z$ relationship (Figure~\ref{fig:fescz}) and the 
predictions at this redshift based upon the \fesclya--\ebv\ relationship 
(Figure~\ref{fig:fesczann}). 
The $z=7.7$ point formally takes the value of \fesclya=0, and is presented with an 
extremely conservative error that is likely to be grossly overestimated
(see \S~\ref{sect:comp}).  
In comparison to $z=5.7$, \fesclya\ has declined by a factor of at least 2 by $z=7$.
We have so far attributed the increase in \fesclya\ to an evolution in 
the dust content of galaxies, and it would be an extravagant departure 
from this evolutionary trend were ISM evolution to suddenly cause a sharp drop 
in \fesclya\ at $z>6$. Several other mechanisms are, however, naturally able to
explain this break in the trend.

\subsubsection{Leaking ionizing radiation}

As discussed previously and by, for example, \citet{Bunker2010} and 
\citet{Bouwens2009b}, the LyC escape fraction at $z\sim 8$ must have been 
around 20--50~\% in order to reionize the universe, depending upon the 
clumping factor of neutral hydrogen. 
Thus, as the galaxy population embedded in the reionization epoch 
evolves into the population observed at lower redshifts ($\approx 3$), 
it must also transition through a phase of modest average \fesclyc\ 
($\approx 0.1$--0.2). At these redshifts, measurements of \fesclyc\ are
emerging that do seem to be consistent with these values 
\citep{Iwata2009,Vanzella2010}, which continue to evolve to lower values with
decreasing redshift \citep[see][]{Siana2010}. 
Furthermore, since at $z\approx 7$ we are looking through the nearest edge of
the reionization epoch into a partially neutral Universe 
\citep[as determined by quasar absorption studies, ][]{Fan2006}, substantial LyC
leakage must occur from the $z\sim 7$ galaxies in order to complete 
reionization. 

If we set \fesclyc$\approx 0$ at $z=5.7$ and hold all
the other properties of the galaxy population constant (i.e. no strong 
evolution of galaxy metallicity or IMF), this estimate of
\fesclyc$\sim 30$\% at $z\approx 7$ would also reduce the nebular emission 
line spectrum to 70\% of its value at $z\approx 6$. This in itself would 
be sufficient to bring the predicted value for
\fesclya\ within $1\sigma$ of the measured value at $z=7$. 
Thus, even in
the redshift 7--8 domain we suggest that the the drop in the \lya\ LF could
be attributed to the drainage of ionizing photons.

\subsubsection{Neutralizing the intergalactic medium}

As the IGM shifts from ionized to neutral, \lya\ photons scatter in 
gas that immediately surrounds galaxies 
\citep{MiraldaEscude1998,HaimanSpaans1999}. This is expected to manifest as a 
drop in the observed \lya\ number counts or LF 
\citep{RhoadsMalhotra2001,Hu2002}, that tails much farther into the 
reionization epoch than absorption tests in quasar spectra. 
Previously \citet{MalhotraRhoads2006} and \citet{Kashikawa2006} have used the 
evolution of the \lya\ LF to look for such signatures of a neutral IGM 
transition but found conflicting results. 
However, the raw differential comparison of LFs only tests the ionized 
fraction if the evolution of the underlying galaxy population is understood to 
an equal, or preferably better, level and \citet{Dijkstra2007}
showed that the evolution reported by \citet{Kashikawa2006} can, for example, 
be explained purely by the evolution of the dark-matter halo population. 
In a similar vein to our own analysis, \citet{Stark2010b} have suggested 
the fraction of LBGs showing strong \lya\ emission to be a preferable signature
of cosmic re-ionization to the evolution of the \lya\ LF. 
Further, from the lack of \lya\ line emission in six out of seven $z \sim 7$ 
galaxy candidates \citet{Fontana2010} suggest that an increasingly
neutral IGM is responsible for reversing the observed trend of the
increasing fraction of strong emitters at redshift below $\sim 6$.

By recasting the problem in terms of the \lya\ escape fraction, we remove the 
question of halo evolution from the problem -- any halo mass function evolution
is accounted for by the LBG population that is used to compute \fesclya. The
drop in the \lya\ LF is also reflected by the \fesclya--$z$ diagram, 
quite securely by $z=7$.  If we hold the ISM 
properties and \fesclyc\ constant, we see that between redshift 
6 and 8 we need to suppress $\gtrsim 50$~\% of the \lya\ luminosity. However, what this
means for the neutral gas fraction is much harder to infer since the fraction 
of photons that scatters in the IGM depends on the exact wavelength with which
\lya\ is emitted \citep{Haiman2002,Santos2004b,MalhotraRhoads2004,Verhamme2008,DijkstraWyithe2010}. All we 
can say with reliability is that the average effective optical depth seen by emitted 
\lya\ photons at $z\sim 7$ would be about 1. 

In summary the dip in the observed \lya\ escape fraction beyond a 
redshift of 6 seems to be real and, holding all other galaxy properties 
constant, a loss of around 50~\% of \lya\ photons needs to be accounted for
by $z=7-8$. Current data can be equally well described by the galaxy population 
emitting this fraction of LyC photons, and by \lya\ photons seeing an IGM
optical depth (at the velocity of the emitting galaxy) of around 1. 
Observational discrimination between the two 
scenarios will remain extremely challenging, but basically calls 
for further deep spectroscopic observations of the $z=7-8$ narrowband and 
dropout candidates, most likely requiring extremely large telescopes.

\subsection{Evolution of the dust content of galaxies}\label{sect:dust}

So far we have been taking advantage of the fact that we have measurements
of the dust extinction in our samples of \halpha\ and UV-selected galaxies. We
have used this to infer the intrinsic star-formation rate density of the
populations, and from there calculated \fesclya\ using Equations 
\ref{eq:fesc_def_ha} and \ref{eq:fesc_def_uv}.
These Equations connect the quantity of \fesclya\ with \ebv, via the ratio of 
the \lya- and UV-derived measurements of the star-formation rate density, 
\rhodstar.  However, in \S~\ref{sect:fescebv} we defined an alternative 
relationship between \fesclya\ and \ebv, based upon analyzing individual 
galaxies, where we provide an empirical relationship between these two 
quantities (Equation~\ref{eq:fescebv}) and relate them simply through 
coefficients. 
Thus we have four quantities (\rhodstarlya, \rhodstaruv, \ebv, and \fesclya), 
that are related by the various coefficients discussed in the previous Sections.

In all the previous Sections we have made use of the measured values of \ebv\ 
but instead we could ignore this measurement, and invert the problem: use the 
observed \lya\ and uncorrected UV star-formation rate densities at a given
redshift to estimate \ebv, using Equation~\ref{eq:fescebv} as a closure 
relation. 
Thus, substituting 
Equation~\ref{eq:fescebv} into Equation~\ref{eq:fesc_def_ha}, we can write: 
\begin{equation}
E_{B-V} = \frac{1}{0.4 (k_\lambda - k_{\mathrm{Ly}\alpha})} \times
	\log_{10} \left( \frac{\dot \rho_{\star,\mathrm{Ly}\alpha}^\mathrm{Obs}}
			{\dot \rho_{\star,\mathrm{UV}}^{\mathrm{Obs}} \cdot C_{\mathrm{Ly}\alpha}} \right) 
\label{eq:closure}
\end{equation}

\begin{figure}[t!]
\includegraphics[angle=0,scale=0.55]{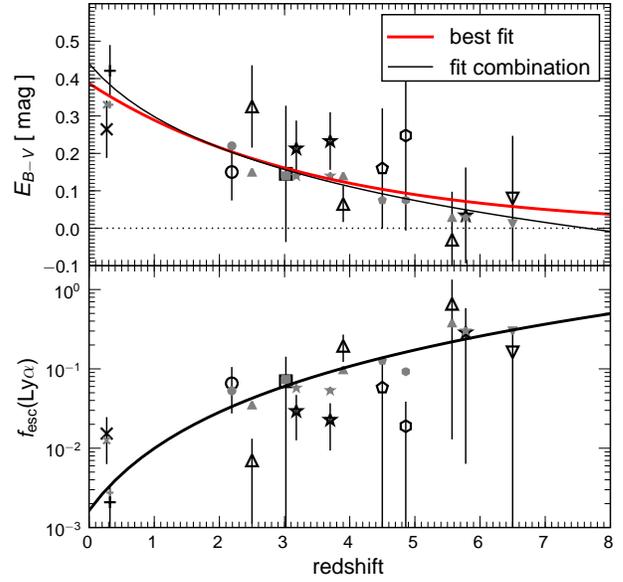}
\caption{{\em Upper}: The evolution of the dust content of galaxies with redshift. 
Black points show \ebv\ derived from the raw observed (i.e. not dust corrected) 
		star-formation 
rate densities in \lya\ and the UV (also \halpha) using Equation~\ref{eq:closure}. 
The gray points show the actual measured values which in general are
well-reproduced by our new method. 
The black lines shows the predictions based
upon the \fesclya--$z$ and \fesclya--\ebv\ relationships derived in
\S~\ref{sect:genres} and \S~\ref{sect:fescebv}, respectively. The red line shows
the best fitting exponential function to these (black) data-points, and is
shown to run slightly flatter, predicting more dust at higher redshifts. 
{\em Lower:}  \ebv\ measurements from the {\em upper} plot
but translated into \fesclya\ using our \fesclya--\ebv\ relationship
(Equation~\ref{eq:fescebv} and Figure~\ref{fig:fescebv}). The gray points
show the same data as Figure~\ref{fig:fescz}, the black line shows
the preferred \fesclya--$z$ power-law. The Figure demonstrates that we would
have arrived at approximately the same conclusions, even if we had no
measurements other than the \lya\ and UV luminosity functions. 
}
\label{fig:dustz}
\end{figure}

Out to $z\approx 6$ we take the data compiled in Table~\ref{tab:fescz},
and compute the observed SFRD from either \halpha\ or the UV, depending on the
redshift. We then use Equation~\ref{eq:closure} to estimate the sample-averaged 
\ebv\ at each redshift, independently of the attenuation measurements 
themselves. In short we ignore the fact that these \ebv\ measurements have been
made, and see if we can recreate them. 
We show the result as black data points in the upper panel of 
Figure~\ref{fig:dustz}, with the actual measurements shown by the the small 
gray symbols.  We then hypothesize that the
dust content of the universe may decrease exponentially, and adopting
a function of the form $E_{B-V}(z) = C_{EBV} \cdot \exp(z / z_{EBV})$, we fit
the coefficients $C_{EBV}=0.386$ and $z_{EBV}=3.42$. Or, the $e-$folding
redshift scale for the \ebv\ evolution is $\approx 3.4$. We show this
relationship in Figure~\ref{fig:dustz} with the thick red line.

By performing this experiment we are throwing away observational information 
and the plot becomes somewhat noisier, but nevertheless it resembles an 
inverted version of Figure~\ref{fig:fescz}. Fundamentally the plot shows a 
decrease in the \ebv\ of galaxies as redshift increases, which is 
consistent with the measurements (Table~\ref{tab:fescz} and gray points). 
This decrease in the dust content of galaxies with redshift is already
much-discussed in the literature for LBGs at $z \sim 2-7$, based upon a gradual
bluening of the UV slopes \citep[e.g.][]{Hathi2008,Bouwens2009a}.
At higher redshift, there is however a tendency for our new method to estimate
higher \ebv\ compared to the measurements obtained directly from the UV
stellar continuum. In the upper panel of Figure~\ref{fig:dustz} we
also show the best-fitting relationships derived in the previous Sections
(black solid line), where we take the redshift evolution of \fesclya\ and 
use Equation~\ref{eq:fescebv} to convert to \ebv\ using our best-fit 
coefficients -- naturally this line almost perfectly reproduces the gray 
points.  

The red line (fit to these data) runs slightly flatter than the black one
(combined fits from the previous sections) and suggests a slightly higher 
\ebv, and and therefore dust content, than measured at the highest redshifts in 
\citet{Bouwens2009a}. 
At $z\sim3$--6 it runs lower, however, than the measurements of 
\citet{Hathi2008} who obtain slightly higher dust attenuations from LBG samples. 

It is interesting to further investigate how the dust obscurations we derive compare 
with other estimates. The SPH modelling of 
\citet{Nagamine2008} and \citet{Dayal2009} already discussed in 
\S~\ref{sect:fesc_comp} 
both predict higher dust attenuations than measured in the $z=6$ dropout 
populations at \ebv=0.15. 
Detailed SED modeling of $z \sim$ 6--8 galaxies by \citet{SchaererDeBarros2010}
also suggest the presence of dust in some high-$z$ LBGs.
Here we estimate \ebv$\approx 0.08$ based upon the new
methodology. Similarly the semi-analytical approach developed in 
\citet{Baugh2005} find \ebv$\sim 0.1$ at $z>3$ when examining the LBG 
population, which is certainly compatible with our estimates in the redshift
3--5 domain.

Since the empirical relationship derived between \fesclya\ and \ebv\ relates
the two quantities directly, for completeness we convert our \ebv--redshift 
estimates to \fesclya\ through Equation~\ref{eq:fescebv}. This enables us to 
approximately re-create the main observational result of this article,
Figure~\ref{fig:fescz}, which we show in the lower panel of 
Figure~\ref{fig:dustz}. Here we show the \fesclya\ estimates derived in this 
Section with black shapes, with the original points from 
Figure~\ref{fig:fescz} shown in gray. In short the difference between the two
sets of points is that in the gray ones the measured dust attenuation has been
applied to the UV star-formation rate density in the computation of \fesclya\
whereas in the black points, this quantity has been estimated directly from the 
observed star-formation rate densities, using Equation~\ref{eq:closure}.
As with Figure~\ref{fig:fescz} this shows \fesclya\ increasing with 
redshift, but the actual estimates of the dust attenuation in the
individual samples have not been used in the derivation of this Figure. The 
overall trend of Figure~\ref{fig:fescz} is maintained, although significant scatter
has been added to the plot. It shows
that even were no \ebv\ measurements available, our main result would have taken 
the same form and the overall trend would have been the same.

\section{Summary}\label{sect:summ}

We have compiled fifteen \lya\ luminosity functions from the literature between
redshifts 0 and 8 and integrated them over homogeneous limits to obtain \lya\ 
luminosity
densities. We have performed the same calculations with \halpha\ emitting 
galaxies at $z\lesssim2.3$, and with ultraviolet selected/dropout samples at 
$z>2.3$, together with their extinctions due to dust. We subsequently used these
dust-corrected luminosity densities to estimate the sample-averaged, volumetric
\lya\ escape fraction (\fesclya) as a function of redshift. 
In summary, we show: 
\begin{itemize}
\setlength{\itemsep}{0cm}%
\setlength{\parskip}{0cm}%
\item{That \fesclya\ increases monotonically from the $\lesssim 1$~\% level at 
$z\approx 0$ to around 40~\% by redshift 6. Over this redshift range, the
evolution can be well described by a power-law of the form 
\fesclya$(z) = C\cdot(1+z)^{\xi}$, for which we obtain 
coefficients of 
$C=(1.67_{-0.24}^{+0.53})\times 10^{-3}; \xi= (2.57_{-0.12}^{+0.19})$. This 
relationship predicts that \fesclya\ should reach unity by a redshift of
$z=11.1^{+0.8}_{-0.6}$. 
}

\item{By combining samples of galaxies at redshift 2--3 for which \fesclya\ and
\ebv\ have been computed, we derive a new empirical relationship between these 
quantities. This provides an effective attenuation law for \lya\ photons that
includes not only the effects of dust absorption, but also those of resonance
scattering by neutral hydrogen. This new relationship takes the form 
$f_\mathrm{esc}^{\mathrm{Ly}\alpha}=C_{\mathrm{Ly}\alpha} \cdot 10^{-0.4\cdot
E_{B-V}\cdot k_{\mathrm{Ly}\alpha}}$, where \klya=13.8 and \clya=0.445.
}

\item{By combining our new \fesclya--\ebv\ relationship with the measured dust
content of (UV- and \halpha-selected) samples in our study, we predict how
\fesclya\ should evolve with redshift, making no reference to \lya\
observations. Between redshift 0 and 6.5, we find that this prediction is fully
consistent with our measurements. Thus we are
able to relate the upwards redshift evolution of \fesclya\ to the
general decrease in the dust content of the galaxy population. We discuss other
effects that could mimic this trend but ultimately find all of them to be 
implausible. 
}

\item{Beyond a redshift of 6 we see a drop in \fesclya\ that amounts to a 
factor of 2--4 by redshift 8. As 
has been done previously, we discuss this drop in terms of an increasing 
neutral gas fraction of the intergalactic medium, but now stress that by casting 
the problem as one of \fesclya, we mitigate the question of halo mass evolution
from diagnostic tests of cosmic reionization. We note however, that that drop
in \fesclya\ could also be explained by a volumetric escape of ionizing photons
of \fesclyc$\approx 50$~\%, which has also been implied at $z=7.5$ by recent 
observations. Unfortunately an observational diagnostic test between the two 
scenarios will remain extremely challenging. }

\item{Using the observed trend between \fesclya\ and \ebv\ derived at $z=2$,
we find a relationship between the observed ratio of \lya/UV star-formation
rate densities and the quantity \ebv. We then use the raw measurements of
\rhodstar\ (\lya\ and uncorrected UV) to estimate how the dust content of galaxies 
evolves with redshift. Our result is a general decrease in dust with increasing
redshift, but not as fast a decrease as measured in UV-selected samples. This
decline is well fit by an exponential function of the form 
$E_{B-V}(z) = C_{EBV} \cdot \exp(z / z_{EBV})$, where $C_{EBV}=0.386$ and 
$z_{EBV}=3.42$.
Using this method, the dust contents we derive at $z=3-6$ are consistent with 
those found by semi-analytical and smoothed particle hydrodynamical models of 
galaxy formation.}
\end{itemize}

\acknowledgments
M.H.\ and D.S.\ are supported by the Swiss National Science Foundation.
G.\"O. is a Swedish Royal Academy of Sciences research fellow supported by the
Knut and Alice Wallenberg foundation, and also acknowledges support from the 
Swedish research council (VR). J.M.M.-H. is funded by Spanish MICINN grants 
CSD2006-00070 (CONSOLIDER GTC) and AYA2007-67965. We thank Mark Dijkstra
and Mauro Giavalisco for useful feedback on the manuscript.

\clearpage

\end{document}